\documentclass[sigplan,nonacm]{acmart}
\usepackage{xspace}
\usepackage{tikz}
\usepackage{amsmath}
\usepackage{subcaption}
\usepackage{tabularx}
\usepackage{algorithm}
\usepackage{algpseudocode}
\usepackage{balance}
\usepackage{printlen}
\usepackage{graphicx}
\usepackage{mwe}
\usepackage{balance}
\usepackage[most]{tcolorbox} 
\usepackage{xcolor}
\title{\sysname: Photonic circuit-switched collective communication for distributed ML}

\providecommand{\vs}{vs. }
\providecommand{\ie}{\emph{i.e.,} }
\providecommand{\eg}{\emph{e.g.,} }

\providecommand{\myparab}[1]{\vspace{1pt}\noindent\textbf{#1} }

\def\shownotes{0}   	

\ifnum\shownotes=1
\newcommand{\authnote}[2]{{ $\ll$\textsf{\footnotesize #1 notes: #2}$\gg$}}
\else
\newcommand{\authnote}[2]{}
\fi

\newcommand{\Anote}[1]{{\color{red}{\bf{\authnote{Abhishek}{#1}}}}}
\newcommand{\Arjun}[1]{{\color{orange}{\bf{\authnote{Arjun}{#1}}}}}

\newcommand{\cutTxt}[1]{\color{black}{}}

\providecommand{\sysname}{\textsc{PCCL}\xspace}
\providecommand{\passage}{\textsc{Passage}\xspace}
\providecommand{\alltoall}{\textsc{AllToAll}\xspace}
\providecommand{\allreduce}{\textsc{AllReduce}\xspace}
\providecommand{\allgather}{\textsc{AllGather}\xspace}
\providecommand{\ltol}{\textsc{Peer-to-peer}\xspace}
\providecommand{\reducescatter}{\textsc{ReduceScatter}\xspace}

\usepackage{titlesec}

\titlespacing*{\section}
{0pt}{0.5ex plus 1ex minus .2ex}{0.3ex plus .2ex}

\titlespacing*{\subsection}
{0pt}{0.5ex plus 1ex minus .2ex}{0.3ex plus .2ex}

\usepackage{caption}
\usepackage{enumitem}

\newlist{compactitem}{itemize}{1}
\setlist[compactitem,1]{label=\textbullet, left=0pt, itemsep=1pt, topsep=1pt, parsep=0pt, partopsep=0pt}

\newtcolorbox{takeaway}[1][]{%
    colback=gray!10,    
    colframe=black,     
    coltext=black,      
    boxrule=0.8pt,      
    arc=3mm,            
    left=4pt,right=4pt, 
    top=0pt,bottom=0pt,
    #1                  
}

\author{Abhishek Vijaya Kumar}
\affiliation{%
  \institution{Cornell University}
  \city{Ithaca}
  \state{New York}
  \country{USA}
}

\author{Arjun Devraj}
\affiliation{%
  \institution{Cornell University}
  \city{Ithaca}
  \state{New York}
  \country{USA}
}

\author{Rachee Singh}
\affiliation{%
  \institution{Cornell University}
  \city{Ithaca}
  \state{New York}
  \country{USA}
}

\begin{abstract}
  Modern distributed ML suffers from a fundamental gap between the theoretical and realized performance of collective communication algorithms due to congestion and hop-count induced dilation in practical GPU clusters.
We present \sysname, a Photonic Collective Communication Library that reconfigures the network topology to match the communication patterns of collective algorithms, thereby eliminating congestion and dilation by creating direct, contention-free circuits between communicating GPUs.
Unlike prior approaches that synthesize algorithms for specific network topologies and collectives, \sysname generalizes to any collective primitive and any topology by adapting the network to match each algorithm's communication pattern.
\sysname's key innovation lies in its hardware-agnostic optimization framework that intelligently decides when to reconfigure based on the trade-off between network reconfiguration delay and congestion/dilation costs, making it practical across different optical hardware with varying switching speeds.
Our evaluation demonstrates that PCCL achieves up to 3$\times$ speedup over state-of-the-art algorithms on 128 GPUs across various workloads, buffer sizes, and topologies, translating to a $1.3\times$ speedup in end-to-end training throughput.

\end{abstract}

\begin{document}

\maketitle

 \section{Introduction}
\label{sec:intro}

Modern machine learning (ML) models are too large to fit in the memory of individual GPUs, causing training and inference to be distributed across multiple GPUs. Distributed ML workloads rely on collective communication primitives --- for example, GPUs aggregate gradients via the \allreduce primitive during training~\cite{mpi}. These primitives are implemented by collective communication libraries (CCLs), like Nvidia's Collective Communication Library or NCCL, in the ML software stack~\cite{nccl,rccl}. The efficiency of distributed ML depends on the performance of collective communication among GPUs. Inefficient collective communication leads to GPUs idling while waiting for data or parameters to arrive over the network~\cite{switchML2021,panama2021, ATP2021, bert, dai2019transformer, deeplight}. 

\myparab{Decades of research on collectives.}
The design of efficient algorithms for collective communication has been an active area of research in high-performance computing (HPC) for several decades~\cite{mpi,collaglo}. Even with decades of progress in developing collective algorithms~\cite{heter-now,bhat2003efficient,eco,networkawarempi}, the gap between \emph{theory} and \emph{practice} remains stubborn: algorithms that are optimal in theory, routinely underperform on practical GPU deployments~\cite{taccl} since these algorithms disregard runtime properties of the underlying topology, including bandwidth variations due to \emph{congestion} and latency variations due to high hop count or \emph{dilation}~\cite{makespan1, makespan2}. This gap has led researchers to synthesize collective algorithms custom-fit to the underlying interconnect topology~\cite{taccl,tacos,sccl,wang2018blink,themis,teccl}. However, synthesis is computationally intractable and does not scale to larger GPU deployments, leading to the use of heuristics that trade off optimality for synthesis time~\cite{taccl, tacos}.

\begin{figure}[t]
    \centering
    \includegraphics[width=0.8\columnwidth]{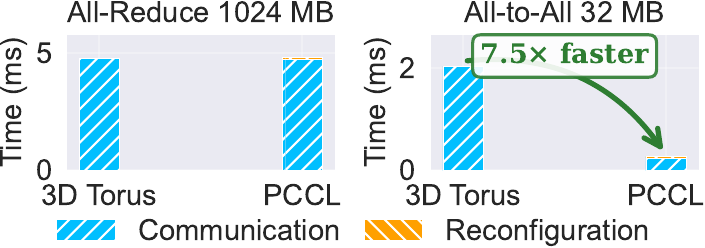}
    \caption{\small Performance of the popular 3D Bucket algorithm for \allreduce collective and Hypercube algorithm for \alltoall collective on a 3D torus topology of size $4\times4\times4$~\cite{googletpuv4}. Note that 3D bucket \allreduce algorithm designed by related perform well on 3D torus and Hypercube is a logarithmic latency algorithm for \alltoall. \sysname matches the performance of 3D torus algorithms for \allreduce and is $7.5\times$ faster than 3D Torus for \alltoall because it does not incur congestion and minimal reconfiguration delay.} 
    \label{fig:motivation}
\end{figure}


\myparab{Collective-aware topologies, not topology-aware collectives.}
We propose a radically different alternative: instead of adapting the algorithm to a fixed network~\cite{taccl,tacos}, we adapt the topology to match the communication pattern of the chosen collective algorithm. We present the \emph{Photonic Collective Communication Library} (\textbf{\sysname}), which reconfigures the network topology to match the communication patterns of known optimal collective algorithms. This ability to reconfigure the network topology is made possible by the adoption of programmable optical fabrics in datacenters for ML~\cite{googletpuv4,tpu-resilience,lightwave,celestialai2024ofc,keren2025,lumorph_ofc}. In each communication round of the algorithm, \sysname can implement direct, contention-free links between exactly the peers that communicate in that round. This eliminates sources of deviation between theory and practice of collective communication algorithms, namely congestion and dilation~\cite{makespan1, makespan2}. Without congestion and dilation, we can reuse decades of work on theoretically optimal collectives (\eg recursive halving/doubling or RHD, ring) and realize their textbook performance in practice.


\myparab{Why is topology reconfiguration key?}
Modern ML workloads expose the fundamental limitation of fixed or direct-connect network topologies: no single collective algorithm can efficiently support the diverse communication patterns within a training iteration. ML workloads invoke thousands of collectives with buffer sizes ranging from kilobytes to gigabytes, each requiring different algorithms for optimal performance. The ring algorithm excels for \allreduce with large buffers but incurs high latency costs for small buffers~\cite{taccl,nccl}, while RHD reduces latency but causes congestion on direct-connect topologies~\cite{swing}. This mismatch is particularly acute in mixture-of-experts models~\cite{moe}, which alternate between latency-sensitive \alltoall operations requiring low-diameter topologies and bandwidth-hungry \allreduce operations that favor ring-like high diameter topologies~\cite{zhao2025efficient}. We show that fixed topologies can force suboptimal performance for different collective algorithms degrading their communication time by 2-4$\times$ due to congestion and dilation. \sysname eliminates this constraint through topology reconfiguration, adapting the network to match each algorithm's communication pattern. Figure~\ref{fig:motivation} shows how this enables using the theoretically optimal algorithm for every scenario---ring for large \allreduce and specialized patterns for \alltoall.

\myparab{Hardware architecture.}
\sysname operates within a datacenter rack-scale GPU deployment, also known as the \emph{scale-up domain}. Electrical switched scale-up domains (\eg NVSwitch) suffer from scaling challenges due to limited switch radix and high power consumption~\cite{ikeda2020large, hotchips34}. To deliver high in-rack bandwidth while keeping energy use in check, ML datacenters are increasingly adopting photonic connectivity within the rack, giving rise to \emph{photonic GPU scale-up domains}~\cite{celestialai2024ofc,hotchips34,lumorph-hotnets,lumorph_ofc,keren2025}. Many architectures have been demonstrated to realize photonic scale-up domains in both industry and academia~\cite{celestialai2024ofc,hotchips34,lumorph-hotnets,lumorph_ofc,karen-ofc23}. While some designs use co-packaged optical transponders~\cite{nvidia_silicon_photonics} to connect GPU servers to a switch, recent work has demonstrated the viability of integrating chip-to-chip photonic connectivity directly into the server board using \emph{optical interposers}~\cite{hotchips34,lumorph_ofc,celestialai2024ofc,lumorph-hotnets}. 

We adopt the interposer-based design as our evaluation platform---not because it is the only viable approach, but because it is an experimentally validated instantiation of a reconfigurable photonic scale-up fabric that enables us to demonstrate \sysname's benefits with realistic hardware parameters (\S\ref{sec:photonic-scaleup}). While direct-connect topologies like 3D Torus~\cite{googletpuv4}, 2D mesh also provide high in-rack bandwidth at a lower power consumption, these topologies are often optimized only for the \allreduce collective and result in suboptimal performance for other communication patterns like \alltoall, as we show in Figure~\ref{fig:motivation}.


\myparab{Technical Challenges.}
Realizing \sysname's vision of topology-adaptive collective communication requires solving three interconnected challenges: (1) \sysname must determine \emph{when} reconfiguration is beneficial. Each topology change incurs a hardware-dependent delay, and \sysname must decide whether this cost is outweighed by the congestion and dilation savings for the remaining communication rounds. (2) \sysname must efficiently map collective algorithms to physical circuits. Given a collective algorithm's communication pattern for each round, \sysname must construct conflict-free optical paths between communicating GPU pairs while respecting hardware constraints: each GPU has limited transmitters and receivers, and each optical waveguide can carry only one circuit per wavelength. (3) \sysname must scale the entire optimization without exponential search time. 



\myparab{\sysname's contributions.}
We address these technical challenges in \sysname, a collective communication library that bridges the theory-practice gap in collective communication by dynamically reconfiguring network topology to match collective algorithms' communication patterns (\S\ref{sec:congestiondilation}). Unlike prior work that synthesizes new algorithms for fixed topologies~\cite{taccl,tacos}, \sysname takes existing, well-understood algorithms and ensures they achieve their theoretical performance (\S\ref{sec:alg}). \sysname makes the following contributions:
\begin{compactitem}
  \item \myparab{Practical performance through intelligent reconfiguration scheduling.} \sysname reduces the reconfiguration decision space from exponential to $O(\log N)$ binary choices by having the collective algorithm dictate the topology for each round, leaving only \emph{when} to reconfigure as the optimization problem. \sysname's optimization can be solved in less than one second for the largest scale-up domains. 

  \item \myparab{Algorithmic flexibility across diverse scenarios.} 
  With microsecond-scale optical reconfiguration, the cost of per-round topology adaptation is far outweighed by achieving consistent theoretical performance across all collective primitives, buffer sizes, and parallelization strategies. This flexibility transforms \sysname from a point optimization into a general solution that works with any initial topology (ring, torus, hypercube) and recovers the theoretical performance of any collective algorithm.

\end{compactitem}

\myparab{Results.}
We show through extensive experiments that \sysname's \allreduce outperforms or matches the performance of well-known optimal algorithms regardless of the initial underlying accelerator topology despite incurring reconfiguration delays. \sysname's \alltoall algorithm outperforms existing well-known algorithms regardless of the initial underlying accelerator topology. Our evaluation shows that \sysname achieves up to $2.5\times$ speedup over state-of-the-art \allreduce algorithms on 128 GPUs, with consistent improvements across different collective primitives, buffer sizes, and scale-up domain sizes. These improvements translate to $1.3\times$ speedup in end-to-end training of popular ML models (\S\ref{sec:endtoend}).

\myparab{Differentiating from work on optical reconfiguration.}
Recent work has proposed leveraging optical reconfiguration in ML datacenters for dealing with accelerator failures~\cite{googletpuv4,tpu-resilience} or for finding rings among accelerators at the start of an entire ML job~\cite{sipml,topoopt}. These systems focus on coarse-grained reconfiguration that occurs either in case of failures or to optimize the initial topology for a specific job. In contrast, \sysname focuses on fine-grained reconfiguration at the scale individual collectives to adapt to the communication patterns of collective algorithms. This difference in timescale and purpose leads to different design choices and challenges, as we will discuss in the paper.

\begin{figure}[h]
    \centering
    \includegraphics[width=0.99\columnwidth]{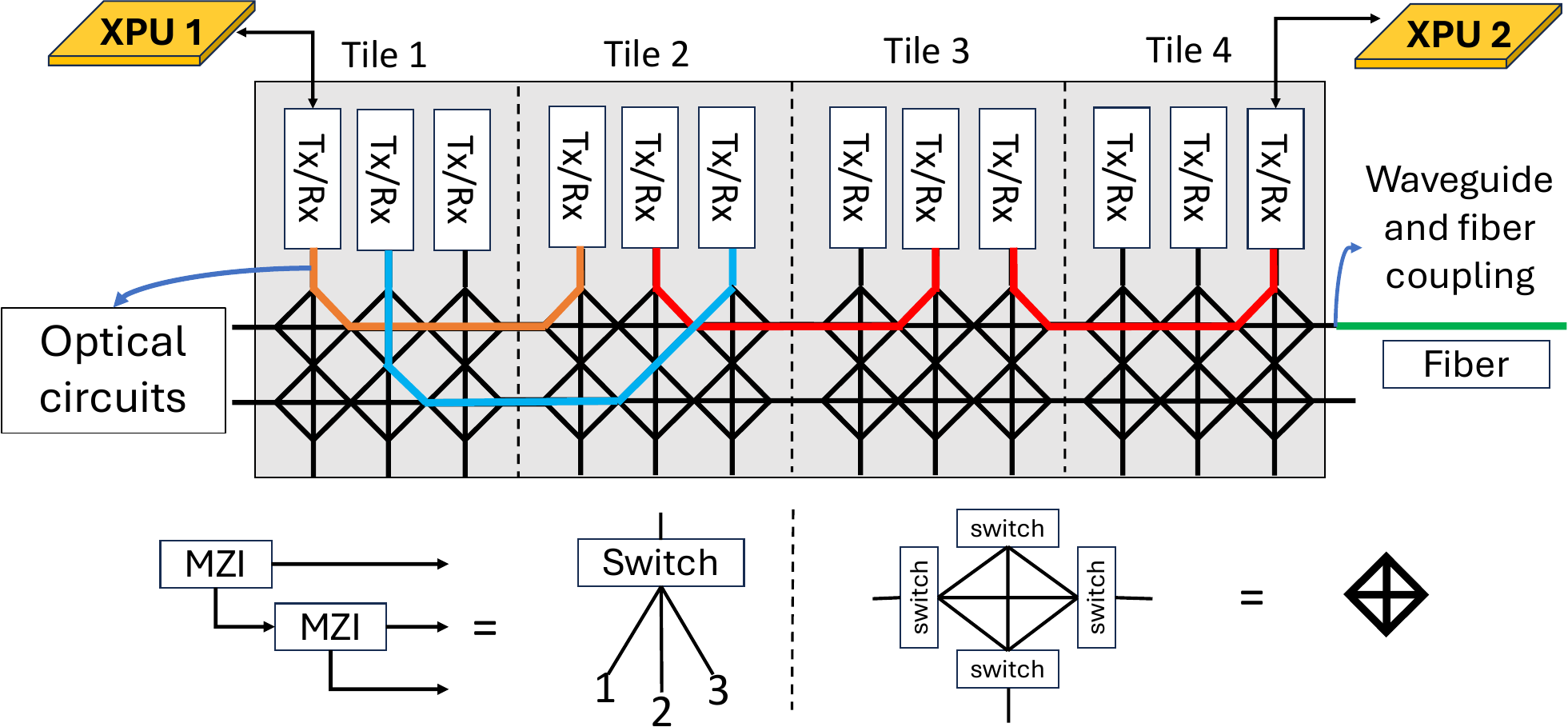}
    \caption{\small Waveguides + Mach-Zehnder interferometer (MZI)~\cite{lumorph_ofc} switches on the interposer form a server-scale optical fabric.} 
    \label{fig:waveguide_grid}
\end{figure}

\section{Photonic scale-up GPU domains}
\label{sec:photonic-scaleup}

Many ML-centric cloud datacenters use optical circuit switches in the scale-out domain to interconnect racks of accelerators~\cite{googletpuv4}. For example, the Google TPU datacenter uses optical circuit switches to interconnect racks of TPUs~\cite{googletpuv4}. Similarly, other cloud providers have replaced or augmented electrical leaf and spine switches of the Clos topology~\cite{vl2} with optical circuit switches~\cite{jupiter-evolving}. Note that these deployments leave the scale-up domains \eg datacenter racks and servers, as monolithic electrical blocks. For example, the NVIDIA DGX~\cite{dgx} GPU servers, NVL72~\cite{nvl72} GPU racks or the Google TPU pods~\cite{googletpuv4}, all use electrical interconnects to network GPUs within a server or a rack.

\myparab{Photonic scale-up GPU domains.}
Recently, the need for higher bandwidth and lower energy consumption in densely connected GPU racks has inspired architectures that introduce optical connectivity even into the scale-up domain, giving rise to \emph{photonic GPU scale-up domains}~\cite{celestialai2024ofc,hotchips34,lumorph-hotnets,lumorph_ofc,keren2025}. Photonic connectivity in scale-up domains aims to replace some of the electrical interconnects within a server or a rack with optical interconnects. Several hardware architectures have been proposed to realize this vision~\cite{celestialai2024ofc,hotchips34,lumorph-hotnets,lumorph_ofc,karen-ofc23}. Some designs use co-packaged optical transponders to connect GPU servers to a silicon photonic switch, as proposed by Nvidia~\cite{nvidia_silicon_photonics}, while others integrate chip-to-chipMZI
 photonic connectivity directly into the server board using optical interposers~\cite{hotchips34,lumorph_ofc,celestialai2024ofc,lumorph-hotnets}.

\myparab{Chip-to-chip optical interconnects in GPU servers.}
In this work, we propose to build scale-up domains with chip-to-chip photonic connectivity within GPU servers. Our design is enabled by newly viable optical interposers like \passage~\cite{lumorph-hotnets,lumorph_ofc}. \passage, an optical interposer, is divided into \emph{tiles} that form a grid. Electrical chips (\eg GPUs, TPUs or XPUs) are bonded on top of the tiles of \passage. These tiles connect to each other using an optical fabric consisting of waveguides and Mach-Zehnder interferometer (MZIs)-based optical switches. Figure~\ref{fig:waveguide_grid} shows the logical layout of the optical fabric on \passage. The I/O (SerDes) ports of an electrical chip connect to transceiver (Tx/Rx) blocks available on \passage tiles. The interconnect fabric on the interposer connects transceiver blocks to each other.

\myparab{Programmable optical fabric.}
A key feature of the optical interposer-based design is its use of \emph{programmable} optical switches in the fabric between GPUs (Figure~\ref{fig:waveguide_grid}, bottom). These switches can be programmed to establish optical circuits between pairs of GPUs. This programmability enables us to adapt the interconnect to changing workloads and communication patterns. In contrast with packet-switched connections on state-of-the-art electrical scale-up interconnects (\eg Nvlinks~\cite{nvlink}, PCIe~\cite{pcie}), optical circuits eliminate contention for network resources on the path between two chips on the interposer. Recent work has shown that optical switches on the interposer can be reconfigured within 3.7 microseconds~\cite{lumorph-hotnets}, making it possible to quickly establish optical circuits between any pair of chips bonded on it.

\myparab{Optical circuits in scale-up domains.}
Figure~\ref{fig:waveguide_grid} shows an example of optical circuits constructed on the interposer by programming MZIs to the appropriate configuration.  Specifically, this configuration makes two unidirectional circuits between tiles $1$ and $2$: one from tiles 1 to 2 (blue) and vice versa (orange). The reticle stitch loss --- loss incurred when the optical circuit crosses a tile --- is $0.2$ dB on average in the optical interposer~\cite{hotchips34,lumorph-hotnets}, allowing us to establish high-bandwidth optical circuits between GPUs. The optical interposer can support multiple such circuits simultaneously, enabling multiple pairs of GPUs to communicate at the same time without contention. This is achieved by the fabric consisting of thousands of waveguides since each waveguide is only a few micrometers wide. 

\begin{figure}[h]
    \centering
    \includegraphics[width=0.65\columnwidth]{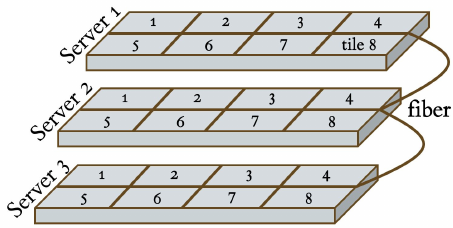}
    \caption{\small Multiple servers cascaded to build a larger cluster.} 
    \label{fig:server_grid}
\end{figure}

\myparab{Cascading optical connectivity in the rack.}
Similar to proposals in previous work~\cite{lumorph_ofc}, servers with optical interposers connect to each other using fibers attached to the waveguides at the edge of a server as shown in (Figure~\ref{fig:waveguide_grid}, right). This allows cascading multiple servers, each with its own optical interposer, to form a larger photonic scale-up domain as shown in Figure~\ref{fig:server_grid}. Within each photonic scale-up domain, the programmable switches and low signal loss enable us to establish optical circuits between GPUs on different servers~\cite{hotchips34,lumorph-hotnets}. In such circuits, the optical signal originates at a tile transponder, traverses waveguides on the interposer, followed by optical fibers between servers and finally terminates at the transponder of the destination tile.

\subsection{Implications for distributed ML}
\label{subsec:collective}

In distributed ML training and inference, intermediate parameters of the model at each accelerator are accumulated, reduced and transferred over the network between accelerators using \emph{collective communication} primitives~\cite{mpi} like \allreduce and \alltoall. This places collective communication on the critical path of both training and inference. Recently, researchers have shown that during model training, accelerators remain idle for large percentages of the time while waiting for inter-accelerator communication to complete~\cite{switchML2021, panama2021, ATP2021, dai2019transformer, deeplight}, highlighting the importance of efficient collective communication.

\myparab{Topology-aware collective algorithms.}
Algorithms for collectives decide the order and amount of data exchanged between accelerators to implement a primitive. Their design has been an active area of research in high-performance computing (HPC) for decades~\cite{mpi,collaglo}. Most of this research assumes the network between accelerators is ideal --- that it has no topology constraints, no contention, and no bandwidth or latency variations at runtime. Recent work has shown that many of these algorithms perform worse than theoretical expectations in real-world deployments since they do not account for the topology and runtime characteristics of the interconnect between accelerators~\cite{taccl,sccl}. A simple example is the use of the theoretically latency \emph{and} bandwidth-optimal recursive halving and doubling (RHD) algorithm for \allreduce~\cite{collaglo}. Depending on the underlying interconnect topology, simultaneous transfers in RHD can overlap on the same network edge, leading to contention and poor performance~\cite{swing}. Cases like these have led researchers to develop topology-aware collective algorithms custom-fit to the underlying interconnect topology~\cite{taccl,tacos,sccl,wang2018blink,themis}.

\myparab{Our proposal: collective-aware network topologies.}
In this work, we propose a radically different alternative: we adapt the interconnect topology to the communication patterns of known performant collective communication algorithms. This allows us to leverage decades of research in optimal collective algorithms and reconfigure the interconnect topology to match the communication patterns of these algorithms to avoid contention and bandwidth variations. 

\subsection{A playground for collective algorithms}
It is tempting to believe that there exists a single ``silver-bullet'' collective algorithm that is optimal for a given collective operation across all scenarios. If this were true, we could simply construct fixed multi-accelerator topologies that match the assumptions of this algorithm and use it for all collective operations, obviating the need for any topology reconfiguration. However, this is not the case. Each collective algorithm has its strengths and weaknesses, and the choice of algorithm depends on the specific requirements of the collective in terms of data buffer size and the characteristics of the underlying hardware. 

\myparab{Selecting the right algorithm.}
Algorithms are classified as \emph{latency-optimal} or \emph{bandwidth-optimal} based on whether they minimize the latency or maximize the bandwidth of communication, respectively. The choice of algorithm depends both on the characteristics of the underlying interconnect, like its bandwidth and latency, and the size of the data buffer. For example, in a high-bandwidth network with high latency, a latency-optimal algorithm may be more efficient. Further, in data parallel \allreduce, the amount of data transferred in each round can be very high due (\eg several GBs) to large model sizes, making bandwidth-optimal algorithms preferable. However, tensor parallel \allreduce in inference transfer smaller amounts of data (\eg several MBs), making latency-optimal algorithms more preferable. Ideally, the collective communication library (CCL) should allow users to choose an optimal algorithm based on these characteristics.


\myparab{Matching topology to the algorithm.}
A collective algorithm only achieves its theoretical performance underMZI
 assumptions about the underlying interconnect topology. So, to truly enable a ``choice'' of algorithm, the network topology must adapt depending on the chosen algorithm; otherwise, the algorithm will be suboptimal due to runtime effects like congestion and dilation (\S~\ref{sec:congestiondilation}). The programmability of photonic scale-up interconnects achieves this by creating on-demand, contention-free optical circuits between pairs of GPUs. This capability introduces a fundamentally richer set of algorithmic choices where communication algorithms can assume direct, contention-free connectivity between communicating GPUs --- unlocking the use of optimal or near-optimal collective algorithms that otherwise underperform on fixed-topology electrical networks~\cite{taccl,tacos}.


\myparab{Photonic Collective Communication Library (\sysname).}
We introduce the Photonic Collective Communication Library (\sysname), a novel collective communication library designed to leverage the dynamic topology reconfiguration capabilities of photonic interconnects in multi-GPU scale-up systems. \sysname unlocks an expansive landscape of algorithmic choices, enabling GPUs to communicate efficiently with dynamically changing sets of peers without incurring contention. By creating on-demand, contention-free optical circuits that precisely match the communication patterns of theoretically optimal collective algorithms, \sysname ensures high-performance collective operations without incurring congestion overheads at runtime.  This flexibility bridges the longstanding gap between theoretical optimality and practical efficiency of collective algorithms, offering distributed ML programmers an unprecedented ability to select collective algorithms tailored to their workload. For example, we can create optical circuits to match the communication patterns of the RHD algorithm for \allreduce, eliminating network contention that would otherwise degrade performance.

\begin{figure}[h!]
    \centering
    \includegraphics[width=0.71\columnwidth]{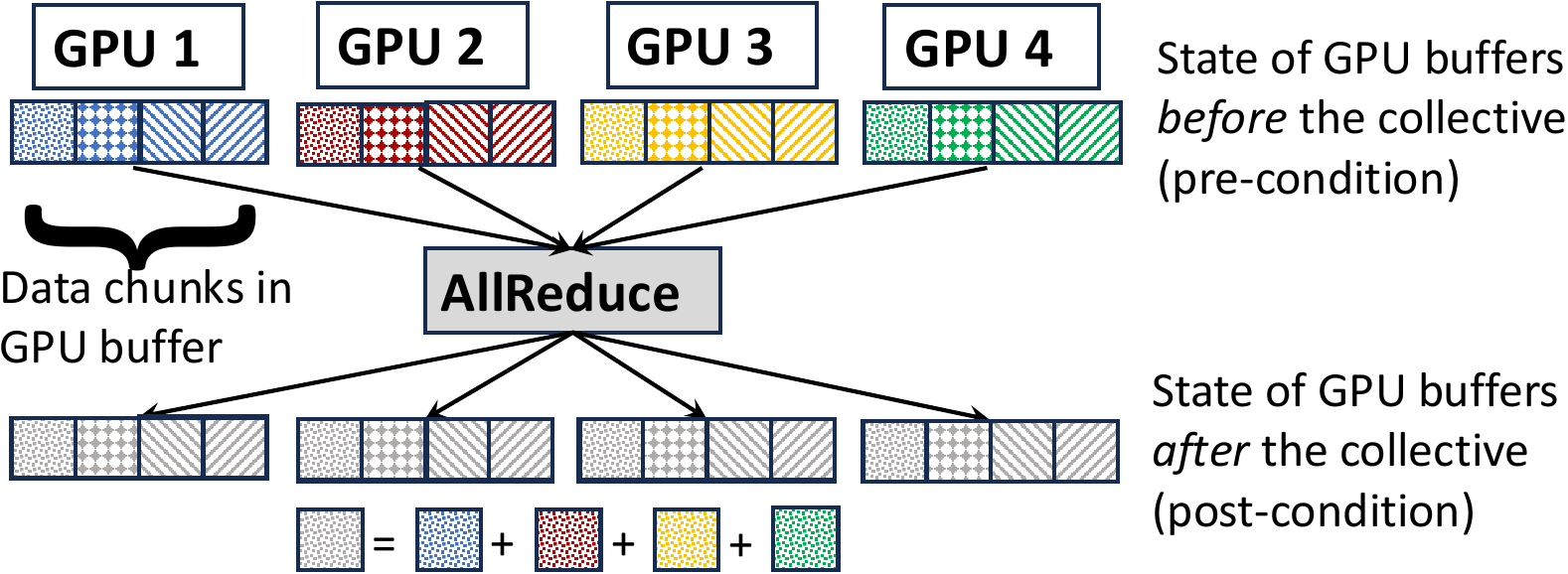}
    \caption{\small describes the \allreduce collective primitive.} 
    \label{fig:ar}
\end{figure}

\section{Congestion + dilation in collectives}
\label{sec:congestiondilation}
A collective communication primitive has a \emph{pre-condition} and a \emph{post-condition} that identifies the data chunks available at each GPU before and after the communication starts and ends, respectively. Figure~\ref{fig:ar} shows the pre-condition and post-condition of the \allreduce primitive. A collective communication algorithm defines the sequence of communication rounds between accelerators that achieve the post-condition of the collective. Each round consists of one or more data transfers between pairs of accelerators. The performance of a collective algorithm depends on the number of rounds, the amount of data transferred in each round, and the characteristics of the communication system. 

We reason about the duration of a collective using the popular $\alpha$-$\beta$~\cite{alphabeta} cost model. Sending a data chunk incurs software overheads of preparing the memory buffer as well as the link latency (\ie propagation delay). Such costs are paid once for every transfer, independent of the size of the chunk. The $\alpha$ constant captures these fixed costs of sending a data chunk on a network link. The $\beta$ costs are a function of the size of the buffer $d$ as well as the bandwidth between the communicating accelerators. So, the cost of sending a chunk of size $d$ from one accelerator to another is $\alpha + \beta \cdot d$, where $\beta=\frac{1}{bandwidth}$. The goal of an optimal collective algorithm is to minimize the total $\alpha-\beta$ cost.

\myparab{$\alpha$-optimal \vs $\beta$-optimal algorithms.}
Decades of HPC research has developed a variety of algorithms for collective communication that optimize the $\alpha-\beta$ costs. Often these algorithms optimize a specific aspect of the communication cost. Bandwidth optimality, or $\beta$-optimality, means that each communicating node transfers the minimum possible amount of data required by the collective communication operation. For example, the Ring \allreduce algorithm between $N$ nodes minimizes the amount of data ($\approx 2 \cdot d$) transferred during the collective schedule, making it $\beta$-optimal but suboptimal in $\alpha$ cost, requiring $2(N-1)$ communication rounds~\cite{nccl-ring}. In contrast, the recursive halving and doubling (RHD) algorithm reduces the number of communication rounds to $2\log n$, improving $\alpha$ costs over the Ring algorithm while still being $\beta$-optimal when $N$ is a power-of-2~\cite{collaglo}.

\begin{figure}[h!]
    \centering
    \includegraphics[width=0.7\columnwidth]{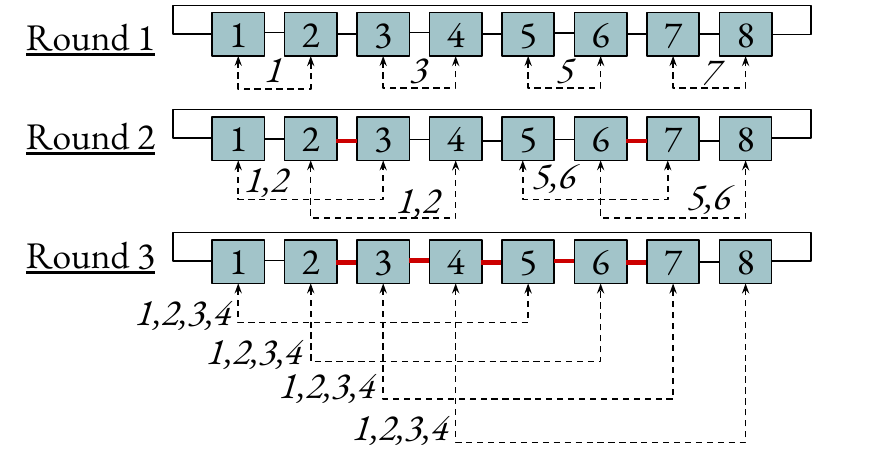}
    \caption{\small shows the optimal communication schedule 
    for \allgather where initially each GPU has one data chunk and the post condition requires that all GPUs have all data chunks ($1-8$). For conciseness, we annotate the arrows with the transfers from the left GPU to the right GPU only, omitting transfers in the reverse direction. This \allgather is the first half of the RHD \allreduce. Note links with overlapping transfers (congested) are shown in red.}
\label{fig:rhd-congestion}
\vspace{-2mm}
\end{figure}


\subsection{Difference between theory and practice}
\begin{figure}[h!]
    \includegraphics[width=0.99\columnwidth]{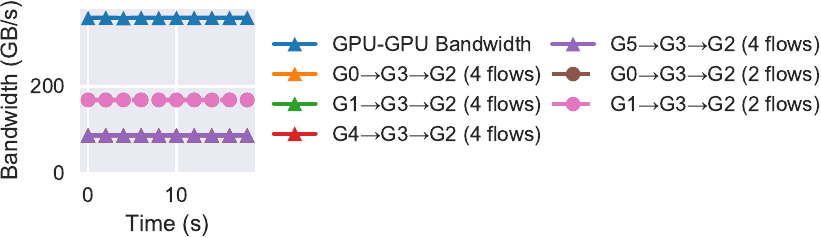}
  \caption{\small Bandwidth between GPUs on different paths ($GX=GPU X$) on a H100 DGX. We copy a stream of 32 MB buffers between GPUs using \texttt{cudaMemCpyAsync}. We store and forward the buffers through GPUs to hop through them.} 
  \label{fig:congestion}
\end{figure}

Despite theoretical guarantees of optimality, the performance of collective communication algorithms in practice is often sub-optimal. In this section we identify the cause for the gap between the theoretical and practical performance of collective communication algorithms. Careful observation suggests that collective algorithms (\eg Ring, RHD) assume that the network is ideal --- each communicating node can send and receive data at the maximum bandwidth of the network link and within each round of communication, all data transfers take the same amount of time. In practice, both these assumptions can be violated due to phenomenon known as \emph{congestion} and \emph{dilation} in the network~\cite{LMR1, LMR2}. We illustrate this with the RHD algorithm for \allreduce:

\begin{compactitem}

\item \myparab{Modeling congestion.}
Consider an example where GPUs participating in an \allreduce are connected physically in a ring topology, also known as a one-dimensional torus~\cite{googletpuv4}. Figure~\ref{fig:rhd-congestion} shows the communication schedule of the RHD algorithm for \allreduce on the ring topology including data transfers in each round. Note that in some rounds, multiple data transfers overlap on the same network link. Overlapping transfers divide the bandwidth of the link among transfers causing \emph{congestion} in the network. Due to this congestion, the effective bandwidth of the link is reduced, increasing the $\beta$ cost of the communication. We model this as a congestion factor $c$ that reduces the effective bandwidth of the link to $\frac{bandwidth}{c}$ where $c$ is the number of overlapping transfers on that link. We experimentally verify $c$ by sharing the GPU2-GPU3 link on a 8 GPU machine across multiple transfers. Figure~\ref{fig:congestion} shows that the bandwidth per transfer reduces by $2\times$ and $4\times$ when 2 and 4 transfers overlap on the GPU2-GPU3 link.

\item \myparab{Modeling dilation.}
In Figure~\ref{fig:rhd-congestion}, communication paths used in each round of the RHD algorithm are not uniform. Some GPUs communicate on a 1-hop path, while others communicate over multiple hops. At each hop, the data needs to be stored and forwarded which introduces some latency. The longer the path length, the higher the latency of sending data between the GPUs. This is known as \emph{dilation} in the network. We model this as a dilation factor $d$ as the number of hops between the source and destination GPU. The dilation factor $d$ increases the effective latency of sending a chunk of size $s$ to $d \cdot \alpha$ instead of $\alpha$.

\end{compactitem}

We extend the $\alpha$-$\beta$ model by introducing additional terms that capture the effects of congestion and dilation on the overall communication cost:

\setlength{\abovedisplayskip}{0pt} 
\setlength{\belowdisplayskip}{0pt}
\setlength{\abovedisplayshortskip}{0pt}
\setlength{\belowdisplayshortskip}{0pt}

\begin{equation}
    \text{communication\ cost} = \sum_{i=1}^{R} (c_i \cdot \beta \cdot d_i + d_i \cdot \alpha)
    \label{eq:congestiondilation}
\end{equation}

where $R$ is the number of rounds in the collective communication algorithm, $c_i$ is the congestion factor for round $i$, and $d_i$ is the dilation factor for round $i$.
Note that $c_i$ and $d_i$ are the \emph{maximum} values of $c$ and $d$ across all transfers in round $i$, since \emph{all} transfers must complete before round $i{+}1$ begins.
Equation~\ref{eq:congestiondilation} captures the total communication cost of a collective algorithm in practical environments, accounting for both congestion and dilation in the network at runtime. Even a known optimal algorithm's performance can degrade significantly if the congestion and dilation factors are high. 

\section{Circuit-switched collectives}
\label{sec:alg}

Theoretically optimal collective algorithms have shown limited utility in practice due to the gap between the assumptions of these algorithms and the reality of the underlying interconnects~\cite{taccl,tacos}. Programmable photonic interconnects provide a unique opportunity to bridge this gap by adapting the interconnect topology to the communication patterns of known performant collective algorithms. We discuss how \sysname implements this vision to achieve optimal collective communication performance in practice.

\myparab{Why can \sysname eliminate congestion in collectives?}
The key to eliminating congestion in \sysname is the use of optical circuits between GPUs. Unlike electrical packet-switched interconnects, which can suffer from congestion when data transfers overlap on the same network link, optical circuits provide contention-free connectivity. Each circuit is a dedicated path between two GPUs, allowing them to communicate without interference from other transfers. This means that all data transfers between GPUs in \sysname occur on direct optical circuits, eliminating congestion delays. 

\myparab{Why can \sysname eliminate dilation in collectives?}
Optical circuits on a photonic interconnect like \passage experience several forms of signal loss, including reticle-stitch loss (0.25 dB~\cite{lumorph-hotnets}), waveguide propagation loss (0.4 dB/cm~\cite{lumorph-hotnets}), and waveguide-to-fiber coupling loss (0.5 dB). To ensure uniform data rates across circuits of varying lengths, the launch power of the laser on the source tile (Figure~\ref{fig:waveguide_grid}) is adjusted based on the length of the path---longer circuits that span more tiles or fibers are provisioned with higher launch power. As a result, circuits in the scale-up domain achieve uniform bandwidth regardless of path length. Thus, when GPUs communicate over optical circuits in the scale-up domain, regardless of the number of hops in the circuit, the data is never stored and forwarded on intermediate GPUs eliminating dilation. While light takes longer to travel on longer paths, the additional time is negligible since it is $0.3\mu{}s$ for every additional 100 meters.

\subsection{Collective-aware topology design with \sysname}
So far we have proposed our vision: \emph{programming optical circuits between GPUs to match the communication patterns of established collective algorithms, thereby eliminating delays due to congestion and dilation}. This flexibility will allow selecting a collective algorithm for the specific characteristics of a workload and achieving its theoretical performance guarantee by adapting the interconnect topology to match the communication patterns of the algorithm. In this section we describe how \sysname implements this vision in practice.


\myparab{To reconfigure or not to reconfigure?}
\sysname considers whether or not to reconfigure the interconnect topology at the start of each round of communication in a collective algorithm, including the first round. The choice of whether to reconfigure the interconnect topology is complex. It depends on the trade-off between two opposing factors: (1) the latency of reconfiguring the interconnect and (2) the potential performance benefits from reconfiguration. As discussed in \S\ref{sec:photonic-scaleup}, programming photonic interconnects incurs a delay that depends on the hardware design. Reconfiguration delays can range from 3.7$\mu$s~\cite{lumorph-hotnets} for \passage to tens of milliseconds for MEMS-based circuit switches~\cite{googletpuv4}. Thus, \sysname must carefully consider if the benefits of reconfiguration, measured by reduced congestion and dilation factors, outweigh the costs.

\myparab{Peeking into future rounds.}
The choice of whether to reconfigure or not is made more complex by the need to make this decision in multiple rounds of the collective algorithm. For example, if \sysname were to reconfigure before the first round of communication, it is entirely possible that the resulting reconfigured topology simply does not allow the communication in the second round of the algorithm (\eg{} the nodes communicating in the second round may not be connected in the reconfigured topology of the first round). Thus, \sysname must also consider \emph{future} implications of reconfiguring in a given round. This requires \sysname to have a global view of the collective algorithm and the communication patterns in all rounds of the algorithm.

\myparab{Isn't topology design intractable?}
At first glance, \sysname's task may seem intractable due to the vast search space of possible topologies and the complex dependencies between rounds. However, \sysname sets up the algorithmic task such that it is not searching for arbitrary topologies by intelligently restricting it to a fixed set of topologies. This significantly reduces the search space and allows \sysname to efficiently compute the optimal sequence of reconfigurations across multiple rounds of communication. Specifically, \sysname formulates this as an optimization problem with the goal of achieving the lowest possible communication cost for a given collective algorithm. 



\begin{algorithm}[t]
    \caption{PCCL Algorithm}
    \textbf{Inputs:} \\
    \begin{tabular}{rl}
        $G_0$                             : & Initial topology \\
        $S = \{G_1, \ldots G_{s-1}\}$: & Set of connected topologies \\
        $R = \{R_0, R_1, \ldots, R_{n-1}\}$: & Set of communication rounds \\
        $R_i \in R$:                          & Set of transfers between GPUs\\
                                        & $(s,d)$ in round $i$ \\
        $I = \{G_i$ $\forall i >= s\}$: & Topologies from connecting\\
                                        & pairs of GPUs in $R_{i-s}$ \\
        $G = (S,I)$                   : & Set of all available topologies\\
        $w_j$:                          & Transfer size for round $j$ \\
        $\alpha, \beta$:                & Cost coefficients \\
        $r$:                            & Reconfiguration delay \\[1 em]
    \end{tabular}
    \textbf{Outputs:}\\
    \begin{tabular}{rll}
        $t_{i,j}$:                      & Binary variable: 1 if round $i$ uses \\
                                        & topology $j, \quad j \in [0 \ldots i+|S|]$ \\
    \end{tabular}\\[1em]
    \textbf{Minimize: } $\sum_{i=0}^{n-1} C_i$\\[1em]
    \textbf{where:} \\
    $C_i = \sum_{j=0}^{i+|S|} t_{i,j} \cdot \text{Cost}(G_j, R_i, w_i, i, j)$ \\[0.5em]
    (\ref{eq:cost}) $\text{Cost}(G_j, R_i, w_i, i, j) = \text{CommCost}(G_j, R_i, w_i) + \text{ReconfCost}(i, j)$ \\[0.5em]
    $\text{CommCost}(G_j, R_i, w_i) = \text{Algorithm}~\ref{alg:alphabeta}(G_j, R_i, w_i, \alpha, \beta)$ \\[0.5em]
    (\ref{eq:reconf_cost}) $\text{ReconfCost}(i, j) = r \cdot \neg\left(\text{Bitmap}(t_{i,j}) \land \text{Bitmap}(t_{i-1,j})\right)$ \\[1em]
    \textbf{subject to}: \\[1mm]
        \begin{tabular}{lll}
            (\ref{eq:onetopo}) & $\sum_{j=0}^{i+|S|} t_{i,j} = 1, \quad \forall i \in \{0, 1, \ldots, n-1\}$ & \\[0.5em]
            (\ref{eq:topodep}) & $t_{i,k} = 1 \Rightarrow t_{j,k} = 1, \quad \forall k \geq |S| \quad \forall j \in [k, i-1]$ & \\[0.5em]
        \end{tabular}
    \label{alg:pccl}
\end{algorithm}

\myparab{Inputs.} 
The algorithm requires multiple inputs. (1) An initial network topology $G_0$, which is the state of the scale-up topology before the collective communication starts. (2) A sequence of communication rounds $R = \{R_0, R_1, \ldots$ $, R_{n-1}\}$, where each $R_i$ is a set of data transfers \((s,d)\) that occur in round $i$ and their corresponding data sizes $w_i$. Note that $R$ defines a collective communication algorithm between $N$ GPUs since it describes which GPUs transfer what data to each other in each communication round. Using $R$ we derive the set $I$ containing topologies resulting from connecting pairs of GPUs in each round's transfers. (3) A candidate set $S$ of standard connected topologies (\eg 2D mesh). We define $G$ as a list of topologies where the first $|S|$ topologies are from set $S$ and the remaining topologies are from $I$. Each topology $G_i \in G$, is a logical topology where an edge represents a direct optical circuit between two GPUs. We describe how \sysname constructs these circuits on the photonic scale-up domain in \S\ref{subsec:circuit}. (4) Latency ($\alpha$) and bandwidth ($\beta$) cost coefficients based on the underlying hardware, and (5) the reconfiguration delay, $r$, of the photonic interconnect.

\myparab{Key decision.}
In each round, the ILP must decide to either reconfigure the topology and incur reconfigure delay or preserve the existing topology and incur congestion/dilation delays. To make the problem tractable, \sysname intelligently restricts the number of topology choices in each round.

\myparab{Managing disconnected graphs.} 
However, restricting the topology choice in each round to only the one that exactly matches the current round's transfer schedule could lead to suboptimal decision-making. The topology derived from communication patterns of a round ($R_i$)
may not connect the set of GPUs that communicate in the next round ($R_{i+1}$), which would require many reconfigurations over subsequent rounds.~\Arjun{Would be best to show a visual example here.} Our key insight is that when the reconfiguration cost is high, we can avoid disconnected graphs by reconfiguring to a standard connected graph (\eg 2D mesh) and incur dilation and congestion costs in the subsequent rounds. \sysname allows for such optimizations by allowing the ILP to pick between (1) reconfiguring to the current round's topology, (2) retaining the existing topology, and (3) reconfiguring to a standard connected graph in $S$.


\myparab{Decision variables and objective.} 
The decision variable is $t_{i,j}$, a binary indicator that equals 1 if round $i$ uses topology $j$. Since each round is allowed to retain the existing topology, the round $i$'s topology can be any of the previous rounds' topologies. Hence, $j \in [0 \ldots |S| + i]$ indexes into the set of all the available topologies for the round: $i$ from the ideal topologies for all previous rounds of the algorithm (including this one) and $|S|$ from the set of standard connected graphs. The algorithm minimizes the total cost across all rounds:
\begin{equation}
\text{Minimize: } \sum_{i=0}^{n-1} C_i
\end{equation}
where $C_i$ represents the total cost for round $i$, computed as
\begin{equation}
C_i = \sum_{j=0}^{i+|S|} t_{i,j} \cdot \text{Cost}(G_j, R_i, w_i, i, j)
\end{equation}
since only the cost of the topology selected for that round is actually incurred.

\myparab{One topology per round constraint.} 
This constraint ensures that for every round $i$, the algorithm selects exactly one topology from the available options:
\begin{equation}
\sum_{j=0}^{i+|S|} t_{i,j} = 1, \quad \forall i \in \{0, 1, \ldots, n-1\}
\label{eq:onetopo}
\end{equation}
\myparab{Topology dependency constraint.} 
In each round, \sysname chooses from a growing set of topologies, including both the predefined connected topologies in $S$ and the topologies derived from transfers in each round, until the current round. If a round uses a topology created from a previous round's transfers, all intermediate rounds must also use the same topology because in each round \sysname is allowed only to reconfigure to the current round's topology or a topology from the connected set $S$:
\begin{equation}
t_{i,k} = 1 \Rightarrow t_{j,k} = 1, \quad \forall k \geq |S| \quad \forall j \in [k, i-1]
\label{eq:topodep}
\end{equation}

\myparab{Cost computation.} 
The communication cost in round $i$ has two parts: the cost of transferring data in the current round using topology $G_j$ and the reconfiguration cost if the topology is changed from the previous round:

\begin{equation}
\text{CommCost}(G_j, R_i, w_i) + \text{ReconfCost}(i, j)
\label{eq:cost}
\end{equation}

We compute the communication cost $\text{CommCost}(G_j, R_i, w_i)$ using Algorithm~\ref{alg:alphabeta}, which finds shortest paths for all transfers and calculates a cost based on dilation (longest path) and congestion (most used edge) as $\alpha \cdot \text{dilation} + \beta \cdot \text{congestion} \cdot w$ on the current round's network topology. The reconfiguration cost accounts for topology change:
\begin{equation}
\text{ReconfCost}(i, j) = r \cdot \neg\left(\text{Bitmap}(t_{i,j}) \land \text{Bitmap}(t_{i-1,j})\right)
\label{eq:reconf_cost}
\end{equation}
This means the reconfiguration cost $r$ is incurred only when switching from one topology to another between consecutive rounds. If the same topology is used in both round $i-1$ and round $i$, no reconfiguration cost is applied.

\begin{algorithm}[H]
\begin{algorithmic}[1]
\State $\text{path\_lengths} \gets [\,]$, $\text{edge\_count} \gets \{\}$
       \For{each transfer pair $(s_i, d_i)$ in $R$}
            \If{shortest path $P_i$ from $s_i$ to $d_i$ exists in $G$}
                \State Add $|P_i| - 1$ to \textit{path\_lengths}
                \For{$j = 0$ to $|P_i| - 2$}
                    \State Let \textit{edge} $\gets$ $(P_i[j], P_i[j+1])$
                    \State Increment \textit{edge\_usage[edge]} by $1$
                \EndFor
            \Else
                \State \Return large penalty
            \EndIf
        \EndFor
        \State $\text{dilation} \gets \max(\textit{path\_lengths})$
        \State $\text{congestion} \gets \max(\textit{edge\_usage})$
        \State \Return $\alpha \cdot \text{dilation} + \beta \cdot \text{congestion} \cdot w$
\end{algorithmic}
\caption{Compute Congestion and Dilation Cost}
    \label{alg:alphabeta}
\end{algorithm}
We note that both Algorithm~\ref{alg:pccl} and \ref{alg:alphabeta} are run offline to generate the reconfiguration schedule for a given scale-up domain. Our evaluation shows both these algorithms converge in less than one second.

\subsection{Creating circuits for \sysname}
\label{subsec:circuit}


\sysname requires that all circuits that underpin the connectivity of the chosen topology in a communication round can exist simultaneously. The number of outgoing circuits from a GPU can not exceed the number of optical transmitters (Tx); the number of incoming circuits can not exceed the number of optical receivers (Rx) available on the tile where the GPU is located. We divide the transmitters uniformly across all required connections by a GPU in a round. If the number of connections are higher, we split the round into multiple rounds until all the connections can be accommodated.

Every optical circuit connecting two GPUs needs to traverse the MZI mesh (Figure~\ref{fig:waveguide_grid}) and also traverse fibers between the servers (Figure~\ref{fig:server_grid}). We model the MZI mesh as a graph and develop an algorithm based on Dijkstra's algorithm to find routes that do not overlap on any edge of this graph to preserve signal integrity. We describe this in Algorithm~\ref{alg:mesh_routing} with further details in Appendix~\ref{appendix:passage}. We randomly sample pairs of waveguides on different edges of the mesh and compute routes for each pair using our algorithm. Figure~\ref{fig:mesh_eval} shows that it finds routes in under 2.5 seconds on a $256 \times 256$ grid with 65,000 MZIs, across different numbers of circuits passing through a server.

To connect servers in our design, we attach a fiber between waveguides on the edge of one server and those on the other server (Figure~\ref{fig:server_grid}). Since there are 100s of waveguides within a server, it is operationally infeasible to connect servers by attaching fibers to all the waveguides at the edges. However, limiting this to a single fiber causes contention as there can only be one circuit of a given wavelength on a fiber. We model the server grid as a graph with servers as nodes and fibers as edges, assuming any circuit that enters a server can also exit it using Algorithm~\ref{alg:mesh_routing}. To route all the required circuits, we develop a flow conserving algorithm (Algorithm~\ref{alg:mcf}, Appendix~\ref{subsec:fiber_routing}) that minimizes overlaps on inter-server fibers. On a 64-server grid, the maximum number of fibers needed to support 100 and 512 random circuits is 7 and 31, respectively. The algorithm converged in under 10 seconds in all our experiments. Since communication in distributed ML is predictable and repetitive, we compute these routes offline and reuse across collective invocations.

\Anote{Are there enough waveguides to accommodate all circuits?}
\section{Benchmarking \sysname}
\label{sec:microbenchmarks}

\begin{figure*}[t]
  \centering
  \begin{subfigure}[b]{0.99\textwidth}
    \includegraphics[width=\textwidth]{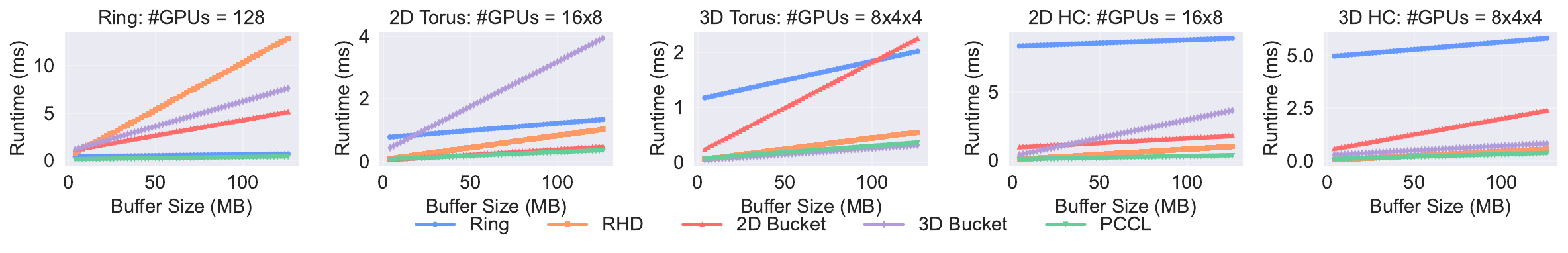}
  \end{subfigure}
  \caption{\small \reducescatter performance across buffer sizes on 128 GPUs with a reconfiguration delay of $5 \mu{}s$. (HC=Grid)} 
  \label{fig:benchmark-1}
\end{figure*}

\begin{figure*}[t]
  \centering
  \begin{subfigure}[b]{0.99\textwidth}
    \includegraphics[width=\textwidth]{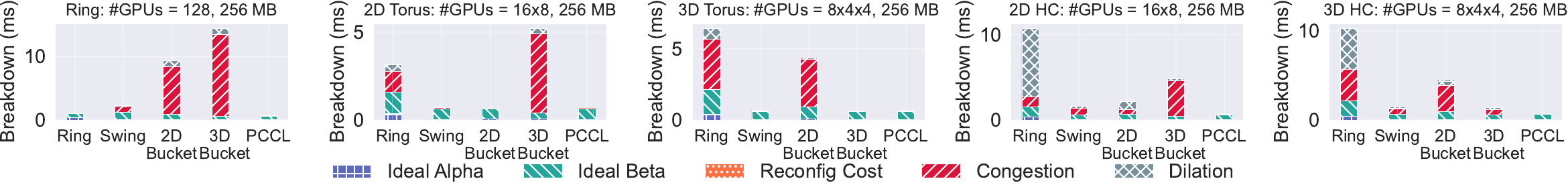}
  \end{subfigure}
  \caption{\small \reducescatter{} performance breakdown for 256 MB buffer on 128 GPUs with a reconfiguration delay of $5 \mu{}s$. (HC=Grid)} 
  \label{fig:cost-1}
\end{figure*}

\begin{figure*}
  \centering
  \begin{subfigure}[b]{0.99\textwidth}
    \includegraphics[width=\textwidth]{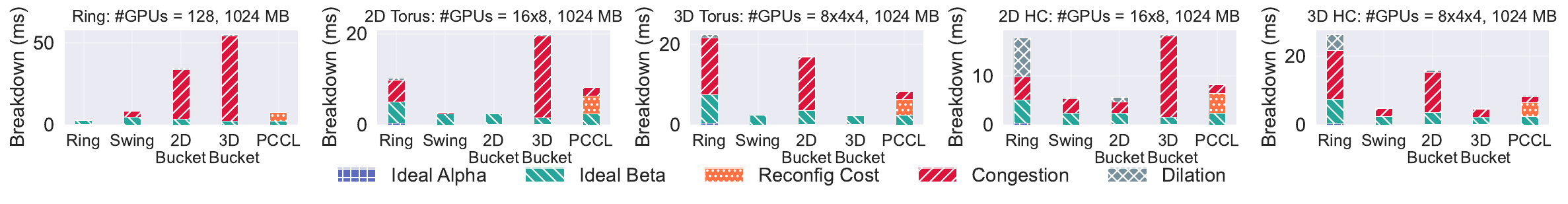}
  \end{subfigure}
  \caption{\small \reducescatter performance breakdown for 1 GB buffer on 128 GPUs with a reconfiguration delay of $1 ms$. (HC=Grid)} 
  \label{fig:cost-2}
\end{figure*}

We evaluate the performance of \sysname's collectives in comparison to the optimal collective algorithms used in popular scale-up direct-connect topologies. These direct-connect topologies have been adopted to provide high in-rack bandwidth with lower power costs and improved scaling compared to switched topologies. We consider 5 baseline topologies: Ring, 2D Torus, 3D Torus, 2D Grid and 3D Grid. Grid is a torus without wrap around links. Many of these topologies are deployed in modern scale-up domains~\cite{googletpuv4, he2025waferllm, tacos}. \textbf{For \sysname{}, these represent the starting topologies $G_0$.}


\myparab{Algorithms.} 
We compare to the hardest baselines: the corresponding known optimal \allreduce algorithm for each topology, also used in practice in state-of-the-art ML scale-up domains. For example, Nvidia's NCCL library~\cite{nccl} implements the bandwidth-optimal Ring algorithm for \reducescatter{}. Similarly, Bucket is the known optimal algorithm for \reducescatter{} on the torus topologies deployed in Google's TPU datacenters~\cite{googletpuv4}. Further, RHD is the bandwidth-optimal algorithm with the lowest known latency cost for \reducescatter{} when the number of GPUs is a power-of-2. We also compare with Swing, which improves RHD on direct-connect topologies~\cite{swing}. \Arjun{Mention what routing algorithm we use when algorithm doesn't match topology}


\myparab{\sysname Inputs.} 
As noted in Algorithm~\ref{alg:pccl}, \sysname{} requires a communication schedule for an algorithm as input. Due to its low latency and bandwidth costs, we provide the communication pattern of RHD as input (\S\ref{sec:congestiondilation}). The interconnect topologies (\eg ring, torus) in \sysname's case are initial configurations which it can adapt across rounds by incurring a reconfiguration delay. We set the reconfiguration delay to 5 $\mu$s, roughly the reconfiguration delay of photonic interconnects like \passage~\cite{lumorph-hotnets}. We use $\alpha = 3\mu{}s$ based on latency measurements of NVIDIA's H100 DGX~\cite{nvidiacudap2p} and $\beta = 1/450$ s/GB since the H100 DGX supports 450 GB/s NVLinks. We evaluate \reducescatter, since \allreduce simply takes $2\times$ the time of a \reducescatter, as it is a \reducescatter followed by a mirror-image \allgather and the two have the same cost~\cite{swing}.
We use the congestion/dilation factors (\S\ref{sec:congestiondilation}) to calculate the extended $\alpha-\beta$ cost using Eq.~\ref{eq:congestiondilation}.

\myparab{\reducescatter performance comparison.}
In Figure~\ref{fig:benchmark-1}, we compare \sysname{} to baseline algorithms while scaling the buffer size for each topology.
Across topologies, \sysname outperforms several baselines, especially at larger buffer sizes on a scale-up domain of 128 GPUs, and matches the ideal algorithm for the topology. Critically, \sysname is the only system that is optimal on all the topologies. All other algorithms incur congestion or dilation on topologies that are not ideal for the algorithm.
Figures~\ref{fig:benchmark-2} and ~\ref{fig:benchmark-3} in the Appendix show the performance of \sysname on scale-up domains of 64 and 32 GPUs, respectively. The performance trends are similar to those on 128 GPUs, with \sysname outperforming all baselines.

\myparab{Communication cost breakdown.}
We unpack why \sysname outperforms baselines by analyzing the communication-time breakdown of each algorithm. 
Figures~\ref{fig:cost-1} and \ref{fig:cost-2} show this breakdown for 128 GPUs with 256 MB and 1 GB buffer sizes and $5 \mu{}s$ and $1 ms$ reconfiguration delays respectively. While baselines suffer from higher dilation delays with the smaller 256 MB buffer, they suffer more from congestion delays with the larger 1 GB buffer. \sysname{} continues to match the best algorithm for each topology when the reconfiguration delay is $5 \mu s$. Furthermore, \sysname is the only algorithm that achieves optimal performance on 2D and 3D grids since they are not an ideal topological fit for any baseline algorithm.

\begin{figure*}
  \centering
  \begin{subfigure}[b]{0.74\textwidth}
    \includegraphics[width=\textwidth]{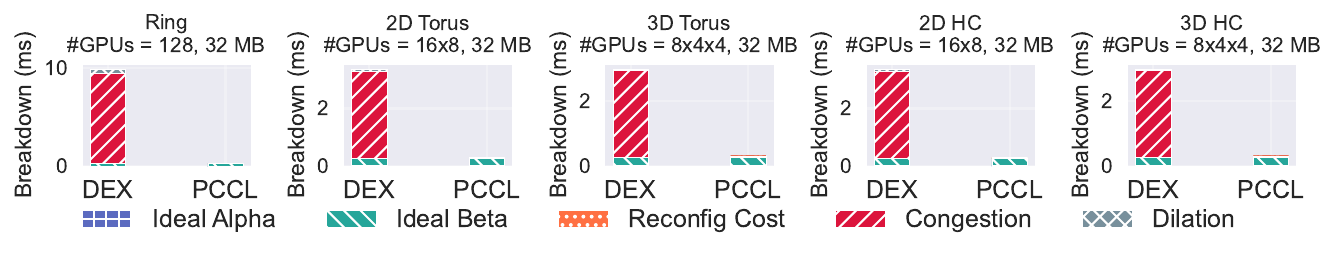}
    \caption{\alltoall{}.} 
    \label{fig:all-to-all}
  \end{subfigure}
  \begin{subfigure}[b]{0.25\textwidth}
    \includegraphics[width=\textwidth]{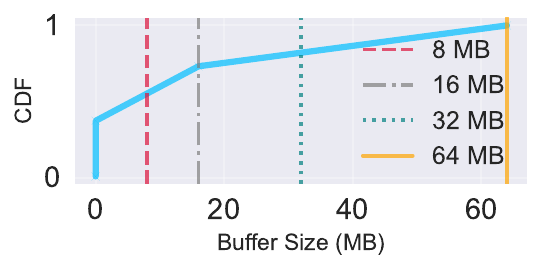}
    \caption{\small All-reduce buffer sizes.} 
    \label{fig:bert_cdf}
  \end{subfigure}
  \vspace{0.2cm}
  \caption{\small shows performance breakdown for 32 MB buffer on 128 GPUs with a reconfiguration delay of $5\mu s$ (HC=Grid). Figure ~\ref{fig:bert_cdf} shows the breakdown of the buffer sizes used in \allreduce calls in Bert.}
\end{figure*}

\myparab{\alltoall performance.}
Scale-up domains must also exhibit strong \alltoall performance due to expert parallelism in MoE models~\cite{moe}. Thus, we compare \sysname to the direct-exchange hypercube \alltoall algorithm (DEX)~\cite{hypercube_alltoall}, a latency-optimal algorithm with $\log_2(N)$ steps suitable for the latency-sensitive \alltoall's in MoE models. \Arjun{Remove this part? that is suitable for small buffers since its bandwidth cost is suboptimal $\frac{m}{2} * \log_2(N)$.} As is standard, we feed DEX's schedule as input to \sysname. Figure~\ref{fig:all-to-all} shows that \sysname outperforms DEX across all the topologies, since none of the topologies are an ideal fit for this algorithm. Our evaluation indicates the \sysname{} outperforms baselines for both \allreduce \emph{and} \alltoall, cementing its optimality and flexibility across use cases.

\myparab{Why does \sysname outperform the baselines?}
In Figure~\ref{fig:cost-1}, where the reconfiguration cost is $5 \mu s$, \sysname achieves the best possible performance on all topologies by adapting the topology to the communication pattern of each round, thereby eliminating congestion and dilation costs. The smaller per-round reconfiguration cost incurred by \sysname in this setting is offset by the savings from eliminating congestion and dilation costs. In contrast, all baseline algorithms incur significant congestion and dilation costs when the topology is not an ideal fit. 
For example, the 3D-bucket algorithm is optimal on a 3D torus topology since it does not incur any congestion or dilation costs. However, it performs significantly worse on other topologies. \sysname's adaptability becomes even more evident in Figure~\ref{fig:cost-2}, where the buffer size is larger and the cost of incurring congestion is higher at 1 millisecond (Equation~\ref{eq:congestiondilation}). In such cases, \sysname can evaluate the tradeoff between reconfiguration costs and congestion/dilation costs and choose whether to adapt the topology accordingly. In Figure~\ref{fig:cost-2}, \sysname chooses to reconfigure only 4 times, instead incurring congestion and dilation, due to the higher cost of reconfiguration. This contrasts with Figure~\ref{fig:cost-1}, where \sysname reconfigures $\log_2(128) = 7$ times.

\begin{takeaway}
\myparab{Key Takeaways:}
\begin{compactitem}
  \item \sysname achieves near theoretical performance of multiple collectives irrespective of input topology.
  \item \sysname intelligently trades off reconfiguration costs with congestion/dilation costs.
\end{compactitem}
\end{takeaway}

\section{End-to-end evaluation}
\label{sec:endtoend}

\begin{figure}[h!]
    \includegraphics[width=0.8\columnwidth]{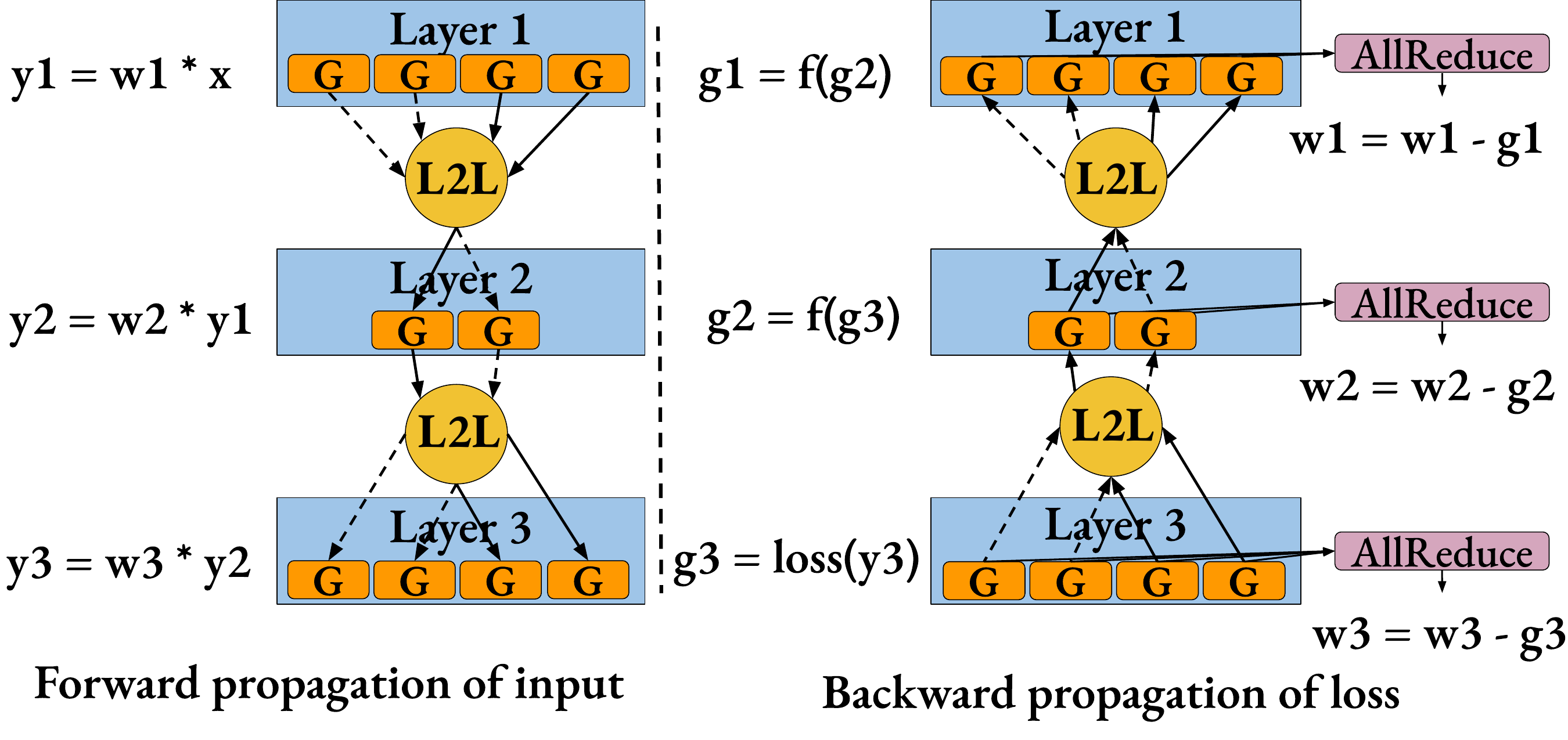}
  \caption{\small Directed graph of a model's iteration. (G = GPU)} 
  \label{fig:taskgraph}
\end{figure}
\balance
In this section, we evaluate \sysname's performance on training transformer models~\cite{bert} using FlexFlow~\cite{topoopt,flexflow,unity}. FlexFlow is the state of the art ML training simulator used to evaluate ML clusters with reconfigurable optical networks~\cite{topoopt}. FlexFlow models distributed training iteration as a task graph (shown in Figure~\ref{fig:taskgraph}) containing compute nodes and communication nodes connected by edges that denote data dependencies. It simulates the end-to-end runtime by iterating through the graph in a topologically sorted order. Prior work using FlexFlow to evaluate optical networks first generates an optimal parallelization strategy using graph optimizations~\cite{topoopt} on the task graph. We retain the same setup and generate an optimal parallelization strategy first and then replace the values of communication nodes with different algorithms to generate associated end-to-end iteration times.


\begin{figure*}[t]
  \centering
  \begin{subfigure}[b]{0.99\textwidth}
    \includegraphics[width=\textwidth]{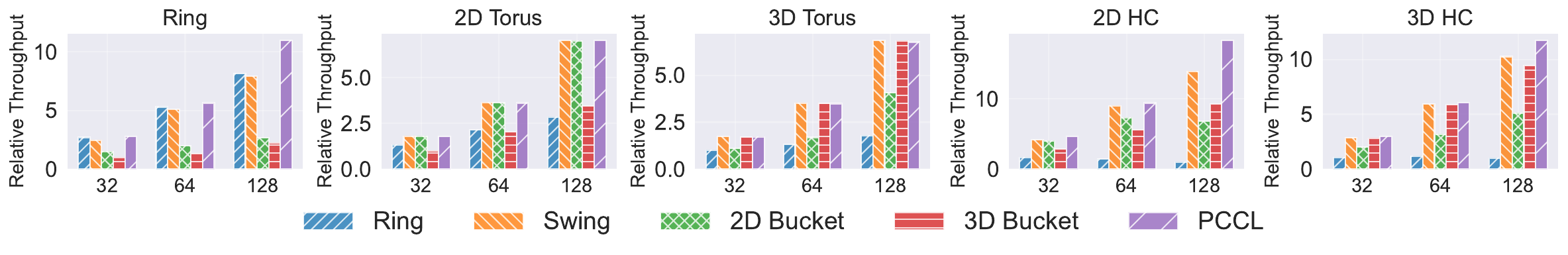}
  \end{subfigure}
  \caption{\small Training throughput of Bert Transformer on 32, 64 and 128 GPUs (HC=Grid). With $5\mu{}s$ reconfiguration delay, \sysname matches the performance of the ideal algorithms on topologies with an ideal \allreduce algorithm (Ring, 2D, 3D Torus). \sysname outperforms all the algorithms on topologies without an ideal algorithm (2D, 3D HC).} 
  \label{fig:end_to_end_throughput}
\end{figure*}

\myparab{Modeling communication nodes.}
The communication in distributed ML training can be broadly classified into two types: \ltol and collectives:
\begin{compactitem}
    \item \ltol traffic arises when a model layer's output shard is consumed by the next layer which is on a different GPU, \eg in pipeline parallelism. This communication pattern may not fit the pattern of a standard collective.

    \item Collectives (\eg \allreduce, \alltoall) are known communication patterns generated by distributed training. For example, data parallelism and tensor parallelism induce \allreduce whereas expert parallelism induces \alltoall.
\end{compactitem}
We tag each communication node in the task graph as one of the above by parsing the communication pattern. For collectives, we feed a collective algorithm schedule to \sysname which reconfigures the topology to optimize collective runtime. For \ltol, \sysname generates point-to-point circuits.

\myparab{Co-scheduling communication operations with \sysname.}
A parallelization strategy can force a layer to enqueue \ltol and \allreduce at nearly the same time, \eg during a backward pass with data and pipeline parallelism. \sysname greedily prioritizes \ltol transfers since they are always on the critical path and cannot be overlapped, whereas \allreduce could, in some cases, be partially overlapped. \sysname implements this within FlexFlow by adding a dependency edge from the \ltol to \allreduce node.

\myparab{Accuracy of large-scale simulation.} 
FlexFlow simulates training on a larger number of GPUs by using a single GPU. It calculates the runtime of compute nodes by executing individual shards of layers assigned to compute nodes on the real GPU. FlexFlow also exposes an interface to calculate communication node costs that integrates with analytical network models like the $\alpha{-}\beta$ model or various network simulators~\cite{topoopt}. In \sysname, we use the $\alpha-\beta$ model---the standard model also used in state-of-the-art distributed ML simulators such as Astra-Sim~\cite{won2023astra} and SimAI~\cite{wang2025simai}---along with congestion and dilation delays, to model the network. Using real execution times for computation nodes along with a well-established, empirically validated model of the network enables accurate simulation of the total execution time.

\myparab{Workload.} 
We simulate the training of a Transformer-based model (12 layers, 16 attention heads, 2048 hidden dimension) across cluster sizes of 32, 64 and 128 GPUs. We execute the FlexFlow simulation on a single A100 GPU with 80 GB of memory. We set the batch size to 16 per GPU and maximum sequence length to 64 tokens. We generate an optimized parallelization strategy to shard the model across all the available GPUs using FlexFlow and then assign different values to the communication nodes corresponding to the collective algorithm and the topology. We set the reconfiguration delay to $5\mu{}s$, modeling the state of the art hardware~\cite{lumorph-hotnets}.

\myparab{\sysname Inputs} 
We use the inputs described in \S~\ref{sec:microbenchmarks} for \sysname. To determine the input schedule for \sysname, we profiled the buffer sizes of the \allreduce nodes in the transformer's training taskgraph. Figure~\ref{fig:bert_cdf} shows that the profiled \allreduce buffer sizes range from 1 MB which is latency sensitive to 64 MB which is bandwidth sensitive. We pick the latency and bandwidth optimal RHD schedule as the input to \sysname to achieve the optimal \allreduce time across buffer sizes. 

\myparab{Baselines.}
We compare \sysname{} with the same baselines from \S\ref{sec:microbenchmarks}. TopoOpt generates ring topologies to reduce congestion across different collective calls. Comparing \sysname with ring algorithm on its ideal topology fit -- ring topology, also serves as the best case comparison with TopoOpt~\cite{topoopt}. 





\myparab{How does \sysname impact ML training?} 
\sysname{'s} performance converges to the best performance on every topology we evaluated and matches the algorithm that is an ideal fit for the topology (\eg 2D bucket on 2D mesh and 3D bucket on 3D torus). Interestingly, \sysname outperforms the ideal algorithm on the ring topology by reducing the $\alpha$ cost from linear to logarithmic in the number of GPUs. The extra reconfiguration cost incurred by \sysname is offset by the improvement in $\alpha$ cost. Swing's~\cite{swing} performance trend is similar to \sysname in matching the ideal algorithm on ring, 2D and 3D torus because it incurs a bounded congestion of $19\%$ and $3\%$ on the 2D and 3D torus respectively. However, \sysname outperforms all the algorithms including Swing on topologies without an ideal \allreduce algorithm (2D, 3D grid) by up to $1.2\times$ since \sysname dynamically creates optical circuits to match the algorithm's communication patterns.

\myparab{How does \sysname handle different reconfiguration delays?}
We simulate the same workload across 32,64 and 128 GPUs with 10, 25, 50 and $500\mu{}s$ as the reconfiguration delay. Figures~\ref{fig:end_to_end_throughput_10}-~\ref{fig:end_to_end_throughput_500} in the Appendix~\ref{appendix:end-to-end} show that at lower reconfiguration delays, \sysname shows similar performance trends as Figure~\ref{fig:end_to_end_throughput}. At higher reconfiguration delays ($500\mu{}s$), instead of reconfiguring in every round, \sysname reduces the reconfiguration and incurs congestion and dilation delays. Thus,~\sysname{} adapts to the reconfiguration delay of the photonic hardware, choosing to reconfigure when the benefits (\ie improved $\alpha-\beta$ costs due to a superior algorithm like RHD, and lack of congestion/dilation from contention-free optical circuits) outweigh the costs (\ie reconfiguration delay).
\section{Related Work}
\label{sec:related}



\myparab{Datacenter-scale optics.}
Google recently replaced spine layer 
packet switches in their datacenter with 
optical circuit switches~\cite{googletpuv4,jupiter-rising,
jupiter-evolving}. Researchers have also used optics for reconfiguring 
the datacenter interconnect~\cite{topoopt,rotornet,firefly,karen4}. 
\cutTxt{Researchers have used silicon photonic interconnects 
for improving ML training~\cite{karen1,karen2,karen3, sipml}.
This work has mainly focused on relatively slow and infrequent 
reconfiguration of the interconnect, called topology engineering.
Unlike them, \sysname actively reconfigures the interconnect even 
within a single communication collective.} 

\myparab{Silicon photonics for ML.}
Recent work has began to use silicon photonic components for machine 
learning workloads~\cite{sipml, karen-ofc23}. Systems like TopoOpt~\cite{topoopt} 
and SipML, use the ring algorithm and reconfigure only once per job. 
TopoOpt finds ring topologies to minimize congestion across collective groups.
SipML\cite{sipml} argues that the reconfiguration cost is justified only 
if the new topology is used for a significant fraction 
of the total circuit establishment and utilization time. 
In contrast, we show that active reconfiguration within a collective
results in end-to-end gains even if 
the utilization time is not significant. In 
\sysname, each GPU sends $log_2(N)$ data chunks 
compared to $N$ chunks in the ring algorithm while retaining the same $\beta$ cost.
Incurring reconfiguration delay $log_2(N)$ times is cheaper than incurring $N \cdot \alpha$
when $\alpha$ and reconfiguration delay are comparable.  


\myparab{Synthesizing collectives.}
Researchers have proposed several algorithms that achieve the theoretical lower bound on the time taken by collectives~\cite{collaglo, makespan1, makespan2}. Recent work~\cite{wang2018blink,cai2021synthesizing} has shown that synthesis is a promising approach for generating collective algorithms for different topologies. However, scaling these approaches to multi-server accelerator topologies has been a challenge. We found that the synthesis time for \allgather and \alltoall  collectives on topologies of two Azure NDv2 nodes and two Nvidia DGX2 nodes using SCCL~\cite{cai2021synthesizing} exceeds 24 hours~\cite{taccl}. \cite{Arzani2023MLFlow} improve on previous work using traffic engineering techniques.
Recent efforts achieve efficient collective communication on heterogeneous 
\emph{networks of workstations}~\cite{heter-now,bhat2003efficient,eco,zhao2024forestcoll,
networkawarempi,sparsecc,sparseallreduce,hoefler-loggopsim,
OptimizationHPC2024,PathOptimization2023,NetworkStatesAware2023,
Collectives2021,COOL2023}.
Most recently, authors in \cite{uw_direct_connect} synthesize efficient direct connect topologies
for collectives in distributed ML. This work targets the scale-out domain 
and evaluates on scale-out network speeds of 100 Gbps while \sysname is targeting
scale-up domains with an order of magnitude higher bandwidth and lower latency.

\label{endOfBody}
\balance
\bibliographystyle{ACM-Reference-Format}
\bibliography{sample.bib}
\newpage
\appendix
\section{End-to-end evaluation experiments.}
\label{appendix:end-to-end}
\begin{figure*}
  \centering
  \begin{subfigure}[b]{0.99\textwidth}
    \includegraphics[width=\textwidth]{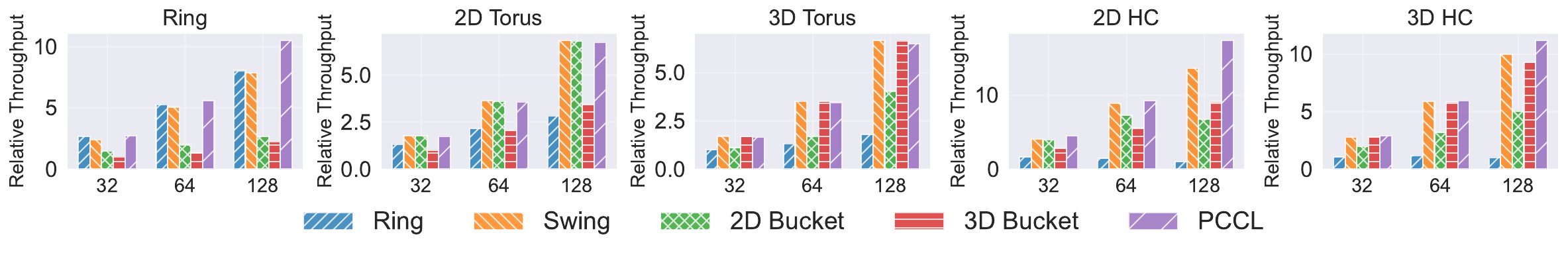}
  \end{subfigure}
  \caption{Reconfiguration delay is $10\mu{}s$. Training throughput of Bert Transformer on 32, 64 and 128 GPUs (HC=Grid).} 
  \label{fig:end_to_end_throughput_10}
\end{figure*}

\begin{figure*}
  \centering
  \begin{subfigure}[b]{0.99\textwidth}
    \includegraphics[width=\textwidth]{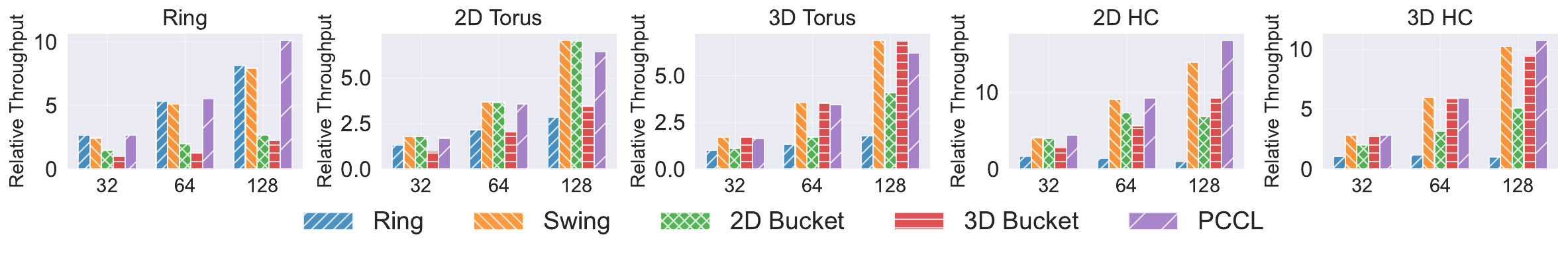}
  \end{subfigure}
  \caption{Reconfiguration delay is $25\mu{}s$. Training throughput of Bert Transformer on 32, 64 and 128 GPUs (HC=Grid).} 
  \label{fig:end_to_end_throughput_25}
\end{figure*}

\begin{figure*}
  \centering
  \begin{subfigure}[b]{0.99\textwidth}
    \includegraphics[width=\textwidth]{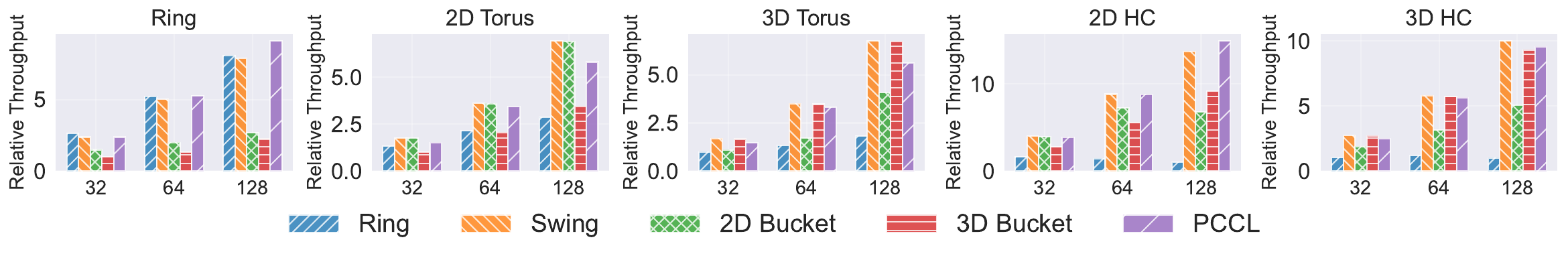}
  \end{subfigure}
  \caption{Reconfiguration delay is $50\mu{}s$. Training throughput of Bert Transformer on 32, 64 and 128 GPUs (HC=Grid).} 
  \label{fig:end_to_end_throughput_50}
\end{figure*}

\begin{figure*}
  \centering
  \begin{subfigure}[b]{0.99\textwidth}
    \includegraphics[width=\textwidth]{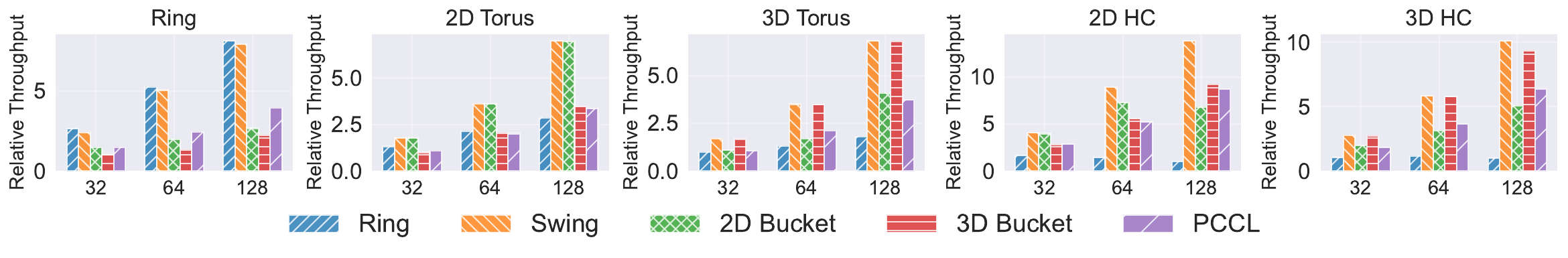}
  \end{subfigure}
  \caption{Reconfiguration delay is $500\mu{}s$. Training throughput of Bert Transformer on 32, 64 and 128 GPUs (HC=Grid).} 
  \label{fig:end_to_end_throughput_500}
\end{figure*}

Figures~\ref{fig:end_to_end_throughput_10},~\ref{fig:end_to_end_throughput_25},\ref{fig:end_to_end_throughput_50},\ref{fig:end_to_end_throughput_500} show the performance of \sysname when the optical reconfiguration delay is 10, 25, 50 and 500$\mu s$ respectively. As the reconfiguration delay increases, the gap between the theoretical optimal performance and \sysname's performance increases since \sysname is forced to pay higher reconfiguration costs or congestion/dilation costs depending on which of the two leads to a lower collective runtime. Instead of always incurring a high reconfiguring cost, \sysname intelligently trades off congestion/dilation for reconfiguration delay whenever it leads to a lower collective runtime. 

\section{Circuits without overlaps}
\label{appendix:passage}

\begin{algorithm}[t]
\caption{Mesh Routing with Edge Reuse Constraint}
\label{alg:mesh_routing}
\begin{algorithmic}[1]
\Require $G=(V,E)$, pairs, MAX\_OVERLAP, PENALIZE\_FACTOR
\Ensure routes, edge\_counts
\State Initialize edge weights $\gets 0$, edge\_counts $\gets 0$
\ForAll{$(s,t)\in$ pairs}
   \For{$k=1 \to \texttt{TRIALS}$}
      \State path $\gets$ ShortestPath($G,s,t,\texttt{weight}$)
      \If{Valid(path, edge\_counts, MAX\_OVERLAP)}
         \State routes[$(s,t)$] $\gets$ path
          \State Update edge\_counts using path
          \State Update weights using edge\_counts
          \State \textbf{break}
      \Else
         \State Penalize overused edges along path
      \EndIf
   \EndFor
\EndFor
\State \Return (routes, edge\_counts)
\end{algorithmic}
\end{algorithm}

We assign routes between source,destination GPU pairs using a shortest-path strategy with edge penalization. We model the MZI mesh as a graph where MZIs are nodes connected by edges which model waveguides. The algorithm maintain a count of how many circuits of a given wavelength are present on a waveguide. A path is valid if no waveguide on the path has overlapping circuits of the same wavelength. When a candidate path is overlapping on certain edges with the same wavelength, the algorithm penalizes that path by multiplying its weight by a penalty factor, making it less likely to appear in subsequent searches. The algorithm greedily tries to pick a valid path by searching for longer paths until a valid path is found. This iterative process balances the path length while ensuring that there are no optical circuits with the same wavelengths on waveguides.



\begin{figure*}
  \centering
  \begin{subfigure}[b]{0.99\textwidth}
    \includegraphics[width=\textwidth]{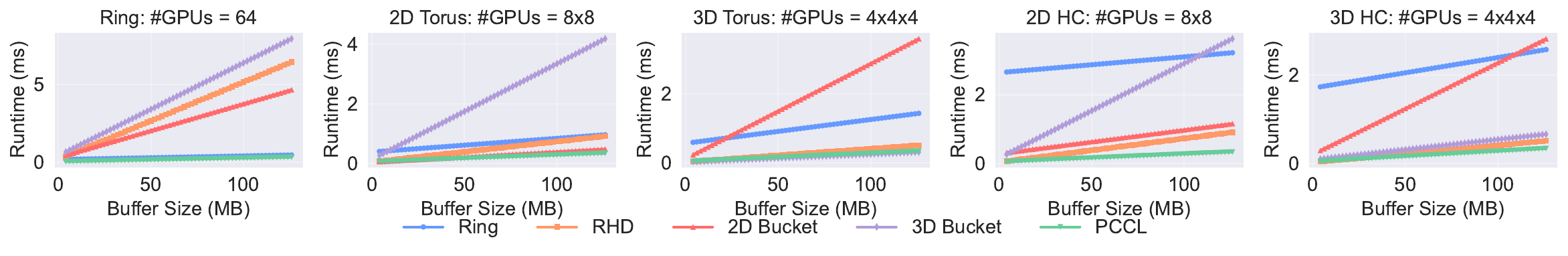}
  \end{subfigure}
  \caption{\sysname \allreduce performance on scale-up size 64, compared to popular collective algorithms and topologies.} 
  \label{fig:benchmark-2}
\end{figure*}

\begin{figure*}
  \centering
  \begin{subfigure}[b]{0.99\textwidth}
    \includegraphics[width=\textwidth]{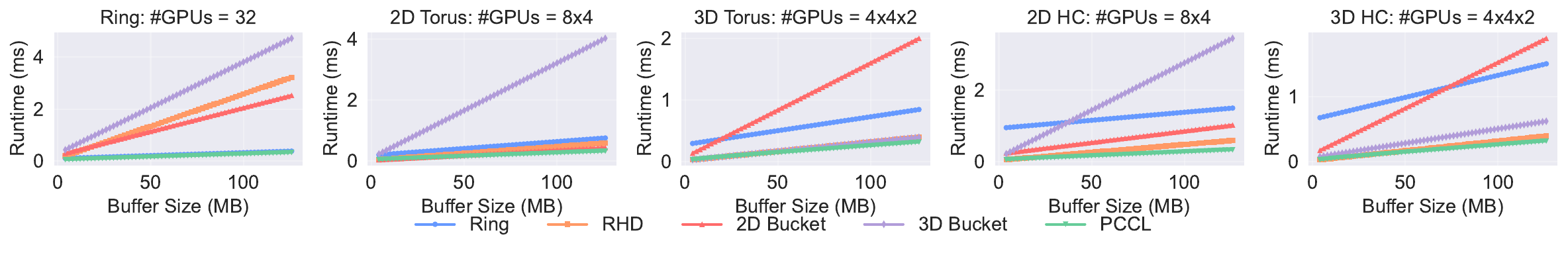}
  \end{subfigure}
  \caption{\sysname \allreduce performance on scale-up size 32, compared to popular collective algorithms and topologies.} 
  \label{fig:benchmark-3}
\end{figure*}

\subsection{Routing to minimize fiber requirement}

\label{subsec:fiber_routing}

\begin{algorithm}[h!]
    \textbf{Inputs:}\\
    \begin{tabular}{rl}
        $G: \langle V, E \rangle $: & $V$ is servers and $E$ is the edges. \\
        $\text{route}_i = (\text{src}_i, \text{dst}_i)$: & $i$-th route request, $\text{src}_i, \text{dst}_i \in V$ \\
        $\text{edge\_count}(u, v)$: & Existing connections on edge $(u, v) \in E$ \\
    \end{tabular}
    
    \textbf{Outputs:}\\
    \begin{tabular}{rl}
        $x_{u,v}^{i} \in \{0, 1\}$: & indicating whether edge  \\
                  & $(u, v) \in E$ is used in a route $i$. \\
        $z$: & integer to minimize overlaps. \\
    \end{tabular}
    
    \textbf{Minimize:} $z$ \\[1em]
    \textbf{subject to} \\[1mm]
    \begin{tabular}{lll}
         & $\sum_{(u, \cdot) \in E} x_{u,v}^{(i)} == 1, \quad \text{if } u = \text{src}_i$ \\
         & $\sum_{(\cdot, u) \in E} x_{v,u}^{(i)} == 0, \quad \text{if } u = \text{src}_i$ \\
         & $\sum_{(\cdot, v) \in E} x_{u,v}^{(i)} == 1, \quad \text{if } v = \text{dst}_i$ \\
         & $\sum_{(v, \cdot) \in E} x_{v,u}^{(i)} == 0, \quad \text{if } v = \text{dst}_i$ \\
         & Flow Conservation Constraint: \\
         & $\forall$ intermediate nodes $v \notin \{\text{src}_i, \text{dst}_i\}$ \\
         & $\sum_{(\cdot, v) \in E} x_{u,v}^{(i)} - \sum_{(v, \cdot) \in E} x_{u,v}^{(i)} = 0$ \\
         & $z \geq \sum_{i=1}^n x_{u,v}^{(i)} + \text{edge\_count}(u, v), \quad \forall (u, v) \in E$ \\
    \end{tabular}
    
    \caption{Path finding with flow conservation.}
    \label{alg:mcf}
\end{algorithm}

Algorithm~\ref{alg:mcf} describes the flow conserving algorithm used to minimize the number of required fibers.It models the scale-up server grid as a graph where nodes are servers and edges are links between the servers. The goal of the algorithm is to find paths for all the input requests while minimizing the overlaps on edges between servers. The value of the objective variable after the optimization is the lowest number of fibers required that can support all the circuit requests.

Note that we can maintain different sets of variables for different wavelengths to accommodate multiple different wavelengths on the same fiber. 

\myparab{Inputs.}
We represent the hardware network as a graph $G = \langle V, E \rangle$, where $V$ denotes the set of servers and $E$ denotes the fiber links between them. For each connection demand $i$, we define a source-destination pair $\text{route}_i = (\text{src}_i, \text{dst}_i)$. Each edge $(u, v)$ has an associated value $\text{edge\_count}(u, v)$, which gives the number of existing connections already occupying that link.

\myparab{Decision variables.}
We define $x_{u,v}^i \in \{0, 1\}$ to indicate whether the path for route $i$ includes edge $(u, v)$. We also define an integer variable $z$ that represents the maximum number of connections present on any edge after including the new routes.

\myparab{Constraints.}
The ILP enforces routing through valid paths using the following constraints.

For each route $i$, we ensure that the source node sends exactly one outgoing flow and receives none:
\[
\sum_{(u, \cdot) \in E} x_{u,v}^i = 1, \quad \sum_{(\cdot, u) \in E} x_{v,u}^i = 0 \quad \text{if } u = \text{src}_i
\]

At the destination, we enforce that the node receives one incoming flow and does not send any further:
\[
\sum_{(\cdot, v) \in E} x_{u,v}^i = 1, \quad \sum_{(v, \cdot) \in E} x_{v,u}^i = 0 \quad \text{if } v = \text{dst}_i
\]

For all intermediate nodes, we conserve flow by balancing incoming and outgoing edges:
\[
\sum_{(\cdot, v) \in E} x_{u,v}^i - \sum_{(v, \cdot) \in E} x_{v,u}^i = 0
\]

We bound the total congestion on the rack by the variable $z$:
\[
z \geq \sum_{(u,v) \in E} \left( \sum_{i=1}^n x_{u,v}^i  \right)
\]

\myparab{Objective.}
We minimize $z$, the maximum number of edges used by connections present across the rack. The main intuition here is the only way to minimize the sum is to minimize every element in the sum. This objective helps avoid having a high number of optical connections on any one edge and minimizes the total additional fibers required for each rack.

\begin{figure*}[h]
  \centering
  \begin{subfigure}[b]{0.3\textwidth}
    \includegraphics[width=\textwidth]{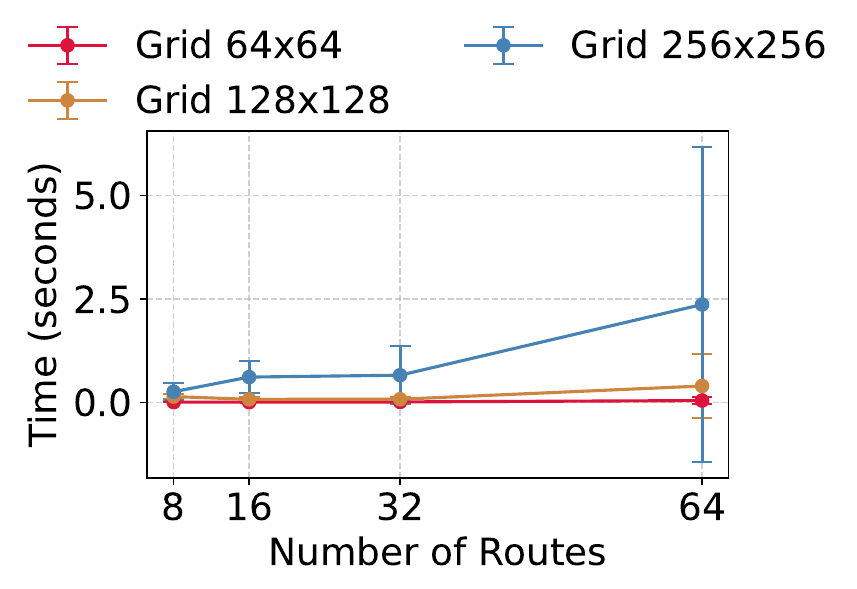}
    \caption{\small{Routes vs time taken.}}
      \label{fig:mesh_eval}
  \end{subfigure} 
  \vspace{2mm}
  \caption{\small{Figure~\ref{fig:mesh_eval} shows the average time taken to compute routes using Algorithm~\ref{alg:mesh_routing} on different mesh sizes.}}
\end{figure*}

\end{document}